# Implementation of Experimental Test Bed to Evaluate Security in Cellular Networks

[1]Nasibeh Mohammadzadeh, [2] Mohsen Hallaj Asghar, [3]N. Raghu Kisore

[1] School of Computer& Information Science,Hyderabad Central University
Hyderabad, Telangana 500046, India
Nasibeh.m@gmail.com

[2] School of Computer& Information Science,Hyderabad Central University
Hyderabad, Telangana 500046, India
Mohsenhallaj62@gmail.com

[3] School of Computer& Information Science,Hyderabad Central University
Hyderabad, Telangana 500046, India
NRaghuKisore@idrbt.ac.in

*Abstract*

The wide development of interconnectivity of cellular networks with the internet network has made them to be vulnerable. This exposure of the cellular networks to internet has increased threats to customer end equipment as well as the carrier infrastructure. If is very difficult to deal with the full fledged infrastructure in a typical service provider network due to operational constraints.However still it's very important for an analysis to evaluate the vulnerabilities and take connective action. This paper introduces the modeling of a small cellular network with security weakness (2G, 2.5G, 3G).This is done by using of the shelf hardware components to set up an experimental test bed. The vulnerability of such cellular networks is attempted with the construction of a ghost GSM tower through shelf hardware components. This paper also discusses concepts of test bed GSM security and possibility of conducting man-in-the-middle attack on the cellular networks model.

## I. INTRODUCTION

Nowadays attackers use GSM prototype towers to simulate attacks and breaking into 2G/2.5/3G networks. Most of the subsystems used are not easy to procure as they are not widely accessible to general public. Almost all of them are owned or regulated by law enforcement agencies. Several hackers also have successfully executed (man-in-the -middle)MITM attacks on cellular networks (including 2G/2.5/3G networks) in the real world. GSM was never designed to be resistant to SIM cloning and eavesdropping but just to serve as a minimallysecure communication medium. GSM uses integrated crypto graphical mechanisms[1]. Previously cryptography had been the domain of military, security agencies, and also banks. Security design of GSM communication aims to provide Authentication, Confidentiality and Anonymity. Authentication is concentrated to ensure that the network operator can verify the identity of the subscriber. In order to do so, it has to be infeasible to clone someone else's mobile phone or SIM card. Authentication is ensured through challenge-response protocol. Confidentiality protects data, voice and sensitive signaling information against eavesdropping on radio path which is done by using symmetric key algorithm to encrypt the radio channel. Anonymity protects against someone who tracks the location of the user or identifies calls either made to or received by eavesdropping on radio path. Rest of this paper is organized as follows: Section II consists of a background and provides technical details of GSM, MITM attacks on cellular networks, and encryption algorithms used at network layers of a GSM network and SDR.Section III is about designing software and implementation. Finally conclusion is represented in thesection IV.

## II. REVIEW OF ESSENTIAL CONCEPTS

### 2. GSM

Global system for mobile communication (GSM) may be a globally accepted normal for digital cellular communication. GSM is that the name of a normalization cluster established in 1982 with the aim of creation of a standard european telecommunications standard that will formulate specifications for a pan-European mobile cellular radio system operative at 900 MHZ.

## 2.1 GSM Architecture

GSM system consists of several sub systems which interact with one another to provide mobile communication service after verification customer credentials. In this sub section, we quickly clarify about some of these subsystems. The three major components of GSM system include: MS (Mobile Station), BSS (Base Station Subsystem), and NSS (Network Switching Subsystem). [2]

**Mobile Station (MS):** It includes the following two components: 1) Mobile Equipment (ME), 2) Mobile Subscriber Identity Module (SIM). Mobile station allows the separation of user mobility from equipment mobility, as well.

**Base Transceiver Station (BTS):** BTS is a piece of equipment that facilitates wireless communication between user equipment and the Telco network. Each cell of mobile network operator has one BTS consisting of high speed directional transmitter and receiver.

**Base Station Controller (BSC):** BSC provides all the management functions and physical links between MSC and BTS. it's a high-capacity switch providing some functions like relinquishing, cell configuration data, and management of frequence (RF) power levels at base transceiver stations. A bunch of BSCs area unit served by associate degree MSC.

**Mobile Switching Centre (MSC):** MSC performs the telephone change functions of the system. It controls calls to and from data systems and different phone. It additionally mobility of subscribers and switches node of a PLMN (Public Land Mobile Network), allocation of Radio Resource (RR), There may be many MSCs in a PLMN.

**Visitor Location Registers (VLR):** Contains temporary information about mobile subscribers that are currently located in MSC service area however whose HLR are elsewhere VLR is answerable for a group of location areas, classically associated with an MSC.

**Home Location Register (HLR):** HLR Contains semi-permanent subscriber information. For all users registered to network, HLR saves users' profiles. MSCs exchange information by HLR. When MS registers into a new GMSC, then HLR sends user's profile to new MSC.

**2.2 Man in the middle attack (MITM):** A MITM attack is a type of attack by eavesdropping on a user's communication with other user(s). This communication is monitored and modified by an unauthorized party. The hacker eavesdrops consistently through interception of a public key message exchange and on the other hand retransmission of message by replacing the requested key by his own.

In these attacks, hackers are primarily targeting specific data about the transactions on computers. This can be anything from an email to a bank transaction that said the hackers begin their investigation of the party of interest. The Universal Mobile Telecommunications System (UMTS) is one of the third generation mobile technologies which are used in most parts of the globe as promotion to existing GSM mobile networks [3].

UMTS subscribers who roam in GSM network are obviously unprotected and unguarded during GSM authentication against false Base Station attacks. Previously, it was widely recognized in available documents of the 3GPP standardization organization [4] that was concentrating on prevention from "false Base Station attacks" for UMTS subscribers roaming in GSM. In fact it recommended modifying the integrity of protection mechanism. [5]. Moreover, the attack described is far beyond of those attacks foreseen by 3GPP in UMTS subscribers. They are even defenseless and exposed to what 3GPP calls a "false Base Station attack", even though subscribers are roaming in a pure UMTS network, and UMTS authentication is used. This attack would be divided into the category as a "roll-back attack". Therefore these attacks utilize and make use of old versions of algorithms and protocols weakness through the use of mechanisms defined to ensure backward similarity and compatibility of newer and stronger versions. Previous results of roll-back attacks consisted of SSL Protocol [6] and encryption mechanisms in GSM.

## III. IMPLEMENTATION

**3. GSM Block Diagram: H**ere we show that model and briefly explain methodology. The diagram shows how a cell phone can connect to internet, laptop or telephone.

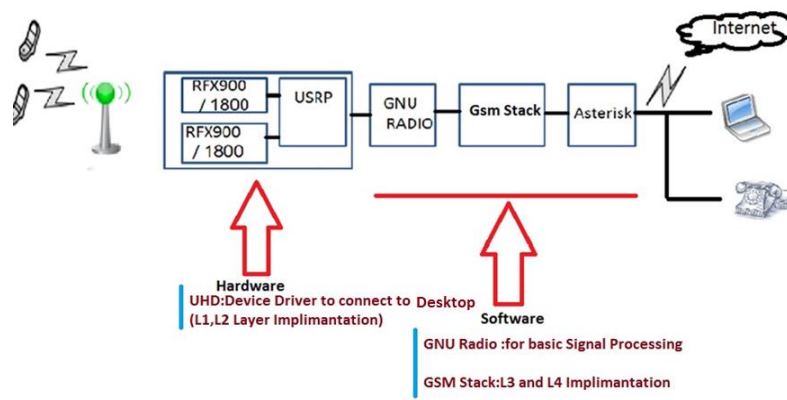

Fig. 3.1: GSM Block Diagram

In fact this is an attempt to provide an open-source Unix application. Generally it uses the Universal Software Radio Peripheral (USRP) to GSM air to cross point and interface the standard GSM handsets. It also applies the Asterisk software PBX to provide connection to calls. It uses the USRP hardware to receive and transmit the signaling that takes place by the usage of GNU Radio framework [6]. The main task of the Asterisk is to interact or interface GSM calls between cellular phones. The GNU Radio is, in fact, a free software development toolkit. The GUN provides signal processing runtime and also processing blocks to implement a software radio which uses readily-available external RF hardware (USRP). Designing USRP (Universal Software Radio Peripheral) is by Ettus Research. It generally serves as a digital baseband and IF section of a radio communication system. It is hardware allowing general purpose computers to function as high bandwidth software radios. A few daughter boards can be possible to apply with the USRP covering from DC to 5.9 GHz. According to the current condition, we might be able to develop and utilize the RFX900 to cover the GSM 850 and 900 bands, or the RFX1800 to cover the GSM 1800 and 1900 bands [7].

**3.1 USRP N210:** Universal Software Radio Peripheral rapidly designs powerful, flexible software radio platforms. In this case we use USRP N210 with three important parts, including Transmit Channel, Receiver Channel and GPS DO (Discipline Oscillator).Transmit Channel and Receive Channel are often used to send and receive data or voice. GPS DO is the combination of GPS receiver and high quality. [8]

**3.2 GNU Radio:** GNU is recursive acronym for GNU's Not Unix. GNU Radio is a platform for experimenting digital communication and also is a software radio construction. GNU Radio is an open source tool. Actually the software is free (Python and C++ Source Code/ Linux Environment). Usually for most of the cases, there is no need for expensive RF test machine.

**3.3 GSM Authentication Process (2G/2.5G) Diagram:**

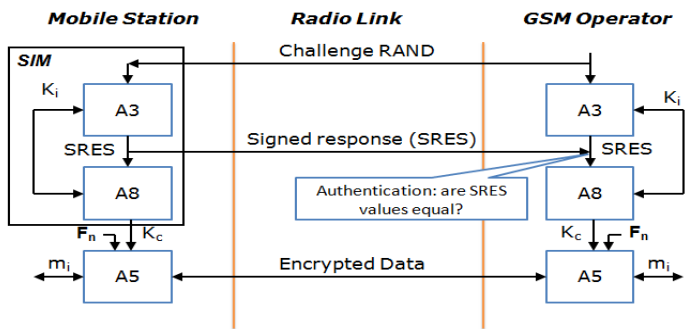

Fig. 3.8: GSM Authentication Diagram

Authentication is based on the SIM (Subscriber Identity Module) which stores individual authentication based key Ki, the user identification International Mobile Subscriber Identity (IMSI), and the algorithm used for authentication A3. The user of specific key Ki is provided by the Telco operator at the time of customer registration.

1. MS sends request to GSM operator, mobile phone also sends its IMSI in a plain text.
2. GSM operator replies with the challenge RAND. Access Control (AC) generates a random number RAND as challenge.
3. MS responds with SRES. SRES is computed as A3ki (RAND) and through RAND Network [VLR] compares results. The operator calculates SRES and compares it with the SRES received from the MS. If they match Kc as shown in the next step.
4. MS and operator calculate Kc=A8 Ki (SRES).
5. Kc is appended by Fn which serves as salt (random data).
6. The encryption of user messages m is done by using A5 algorithm and (Kc||Fn) as the key.

MITM attack can be executed by a ghost Base Station. The ghost Base Station can stand between MS and GSM Operation. Then it can play the role of BS for MS and for GSM Operation as MS. Afterwards it can request BS to turn off encryption and use A5/0. In addition to MITM, the ghost Base Station can also impersonate a genuine MS.

**3.4 Mobile Phone Captured for GSM-900 and GSM-1800 Base Stations:** First of all we need to find out the frequency of channel and play roles as BS. In this case we have four parts: TMSI, IMSI, AGE and used. TMSI is Temporary Mobile Subscriber Identity. It is the identity that is commonly sent between network and cell phone .TMSI is assigned via the VLR, anywhere and to any mobile. IMSI is International Mobile Subscriber Identity that has a particular and unique number to each mobile. Actually AGE and used explained the first time mobile connection to device and also the last time use of mobile to make a call or send a message.

**3.5 Results:** After configuring the power setting in our test bed, software: fedora 15, UHD-003.006, gnuradio 3.7.0, OpenBTS 3.1 also Hardware: USRPN210 + GPSDO from Ettus Research. Installed Asterisk as the soft IP phone. as matter of fact we dont use proprietary SIM Cards, use the regular SIM cards issued by a legal Telco Operator.after a week's struggle able to make a mobile handset connect to the tower. Here is the log for the same.When sender send message via mobile phone ,capture this with usrp2 ,and at first updated sender location and requested to update sender mobile phone's data . At second check International Mobile Subscriber Identity(IMSI) which that include in withelist or not, if (IMSI) has'nt any problems register and allow this to send a message .Third, sender can find which type of message has be send and when sender message delivery confirm about this matter , and mobility massage class send CPack as sending message is ok .Finally we obtained the following results after sending message:

- MobilityManagement.cpp:194:LocationUpdatingController: MM Location Updating Request UpdateType=(7) LAI=(MCC=404 MNC=07 LAC=0xfffe) MobileIdentity=(IMSI=404070536564665) classmark=(revision=1 ES-IND=1 A5/1=0 powerCap=3)
- Jan 20 16:10:58 localhost : INFO 3062872944 16:10:58.2 MobilityManagement.cpp:239:LocationUpdatingController: not checking white-list for IMSI404070536564665
- Jan 20 16:10:58 localhost : INFO 3062872944 16:10:58.2 MobilityManagement.cpp:420:LocationUpdatingController: registration ALLOWED: IMSI=404070536564665
- Jan 20 16:10:58 localhost : INFO 3062872944 16:10:58.9 MobilityManagement.cpp:121:sendWelcomeMessage: sending Control.LUR.OpenRegistration.Message message to handset
- Jan 20 16:10:59 localhost : INFO 3062872944 16:10:59.4 SMSControl.cpp:388:deliverSMSToMS: sending CP-DATA TI=5 RPDU=(01f503a10000004b0403a101f100004110026101 952244d7327bfc6e9741f437888e2e838ed326885e9ed341 ee32fdfe96af5d2050f65d978392cd6912 949e8340c9e634a9a3c168b01bac36b3d56c349bad06)

- Jan 20 16:11:02 localhost : INFO 3062872944 16:11:02.4 SMSControl.cpp:447:deliverSMSToMS: MTSMS: sending CPAck

## IV: Conclusion

In this paper we represented an overview of how to build a ghost Base Station using commercially available off the shelf hardware components and open source software. We also succeeded in demonstrating the captured SMS messages of friendly mobile users to understand the cryptographic algorithms used. We came to know that none of the Telco operators provides encryption at the network layer. Hence we recommend that alike business and banks design their mobile applications where in encryption is done at the application layer.

While the developed cellular Base Station has practical limitations with respect to transmission range and region of coverage, both issues can be overcome by using proper hardware. Our system was built using a 100mwatt power source; therefore it provides us with a transmission range of 150 feet. The transmission range can be easily increased to a much higher level by using an external power amplifier. Power amplifiers of 1Kwatt are available and can be used in our hardware to boost transmission range to 1.5KM. Our prototype uses a directional antenna; hence cannot scan mobile phones in all directions. Though expensive, omni-directional antennas can be easily provided from market.This is just to demonstrate the proof of concept,extension may easily be realized.

[1]**First Author**: Nasibeh Mohammadzadeh, M.Tech Information Security and banking technology (IT) - School Of Computer & Information Science at University of Hyderabad, India(2012-2014)- Interested to security in cellular networks and mobile telecommunication networks- nasibeh.m@gmail.com

[2]**Second Author:** Mohsen Hallaj Asghar, M.Tech Information Security and banking technology (IT) – School of Computer& Information Science at  University of Hyderabad, India(2013-2015),interested  to security in Internet of Things, challenges and application in (IoT) and security in cellular networks and social network in IoT- mohsenhallaj62@gmail.com

[3]**Third Author :** Dr. Raghu Kisore Neelisetti Assistant Professor,IDRBT ,university of Hyderabad  –Hyderabad ,India –Ph.D. in Computer Science from Auburn University, December 2009,Master's Degree in Computer Science from IIT Madras in 2002.
NRaghuKisore@idrbt.ac.in